\documentstyle[PASJadd,epsf]{PASJ95}
\def \H2{H{\cmr 2}}
\def \ds{\displaystyle}
\def \gtrsim{{\;{\ds\lower.7ex\hbox{$>$}\atop\ds\sim}\;}}
\def \lesssim{{\;{\ds\lower.7ex\hbox{$<$}\atop\ds\sim}\;}}

\def \L60{$\nu_{\rm 60}$L$_{{\rm 60}\mu m}$}

\newcommand{\vol}{999}
\newcommand{\no}{1}
\newcommand{\lett}{}
\newcommand{\spage}{1}
\newcommand{\rdate}{1999 June 17}
\newcommand{\adate}{1999 August 26}

\begin{document}
\setcounter{page}{1}

\title{Color and Morphology of Galaxies \\
in the Region of the 3C 324 Clusters at $z \sim 1.2$}

\author{Masaru {\sc Kajisawa} and Toru {\sc Yamada}\\
{\it Astronomical Institute, Tohoku University, Aoba-ku, 
Sendai, Miyagi 980-77} \\
{\it E-mail(MK,TY): kajisawa/yamada@astr.tohoku.ac.jp }}

\abst{
 We investigated the color and morphology of optically selected
galaxies in the region of clusters at $z \sim 1.2$ near to the radio
galaxy 3C 324 using archived data taken with the Hubble Space
Telescope. The faint galaxies selected at the HST {\it F702W} band
that contribute to the surface-density excess of the region have wide
ranges of color, size, and morphology, which are not likely to be due
to contamination by foreground galaxies. Namely, the rest-frame
ultraviolet emission properties of the galaxies in the clusters are
not very homogeneous; various amounts of star-formation activity may
occur in a significant fraction of them. Although our analysis is
purely statistical, we find that typical star-forming galaxies with
blue colors have a relatively late-type morphology compared to the red quiescent population in the systems.
}
\kword{galaxies:~clusters of --- galaxies:~evolution --- galaxies:~formation}

\maketitle
\thispagestyle{headings}

\section{Introduction}

 Recently, more than several clusters and candidates at $z \gtrsim 1$
have been discovered (Dickinson 1995, 1997a,b,c; Yamada et al. 1997;
Stanford et al. 1997; Hall, Green 1998; Ben\i tez et al. 1999;
Tanaka et al. 1999). While most of these objects are selected by the
surface-density excess of the quiescent old galaxy population, whose
characteristic optical--NIR colors are relatively easy to be traced, significant `active' evolution of the galaxies is also naturally expected to be observed in such high-redshift clusters.

 Especially, it is inferred from previous observations that the
star-formation activity in high-redshift clusters is higher than that
in the intermediate-redshift and nearby clusters. At intermediate
redshift,  a fraction of blue or emission-line galaxies is known to
increase rapidly with the redshift (Butcher, Oemler 1978, 1984;
Dressler, Gunn 1982, 1983, 1992; Rakos, Schombert 1995), which may
reflect the evolution of star-formation activity. A simple
extrapolation of this Butcher--Oemler effect gives more than a $50\%$
blue galaxy fraction at $z\sim 1$. There is also a significant
fraction of galaxies with the {\it post}-starburst signature in
intermediate-redshift clusters (e.g., Dressler et al. 1999; Poggianti
et al. 1999), which directly implies that some significant star
formation must have occurred and terminated at some epoch of higher
redshift. This evolution of star-formation activity is probably
related to a morphological transformation of galaxies in clusters
(Dressler et al. 1994, 1997, 1999). It is interesting to study how
these expected evolutional features are really observed in clusters at
$z \gtrsim 1$.

 So far, however, the number of clusters at $z \gtrsim 1$ whose
photometric properties have been studied in details is still very
limited.  Tanaka et al. (1999) recently investigated the colors of the
galaxies in the cluster 1335.8+2820 at $z \sim 1.1$. This cluster has
a very wide distribution of optical and near-infrared colors, which seems to be due to the effect of various amounts of star-formation activities occurring in the cluster galaxies rather than to the age and/or metallicity differences or the contamination by the foreground objects. For galaxies with $K < 19$ mag, the fraction of galaxies which show UV-excess to the quiescent-galaxy models is more than $75\%$, and most massive galaxies thus no longer seem to evolve passively. Then, it is a question whether the cluster 1335.8+2820 is a peculiar example or it shares the general characteristics of high-redshift clusters.

 In this paper, we consider the color and morphology of
optically-selected galaxies in the region of clusters at $z \sim 1.2$
near the radio galaxy 3C 324 using the archived data taken with the
Hubble Space Telescope (HST). The clusters are recognized by Kristian,
Sandage, and Katem (1974), by Spinrad and Djorgovski (1984), and
firmly identified by Dickinson (1995). They have been
spectroscopically confirmed, and have been revealed to be a
superposition of the two systems at $z = 1.15$ and $z = 1.21$ (Dickinson 1997a,b). We describe the data and our procedures to obtain the colors and the morphological parameters of the galaxies in section 2 and show the characteristic properties of the galaxies in the region of the clusters in section 3. In the last section, we give the conclusions and discussions. 

\section{The Data and the Observed Quantities}

 A deep WFPC2 field surrounding the radio galaxy 3C 324 at $z = 1.21$ was
imaged using the {\it F702W} and {\it F450W} filters on 1994 May and June, and
1997 June, respectively. We analyzed the calibrated data of the field
downloaded from the archive of the Space Telescope and Science
Institute. The total exposure time is 64800 s (18.0 hr) with the {\it
F702W} band and 15120 s (4.2 hr) with the {\it F450W} band.
%

 Initial source lists were constructed using the SExtractor software
(Bertin, Arnouts 1996). We set the surface-brightness threshold for a
pixel as $\mu_{\it F702W}$ =26.7 mag arcsec$^{-2}$. If the counts of
more than 12 continuos pixels are above the threshold, they are
recognized as sources. For these sources, pseudo total magnitude
values (MAG BEST from SExtractor) are obtained for the {\it F702W} and
the {\it F450W} filters in STMAG system. STMAG is defined as
$m_{\rm ST} = -21.10-2.5$ ${\rm log}f_\lambda$ where $f_\lambda$ is expressed
in erg cm$^{-2}$ s$^{-1}$ \AA $^{-1}$ (HST Data Handbook Vol.1). The
detection limit is {\it F702W} $\sim 28$ mag and {\it F450W} $\sim 28$
mag, respectively, which corresponds to $R_{\rm AB} \sim27.4$ mag and
$B_{\rm AB} \sim 28.4$ mag, respectively. The uncertainties in magnitude are 0.1--0.3 mag at these detection limits.
 We analyzed the color, half-light radius, mean surface brightness,
and other morphological parameters such as asymmetry and concentration
indices, of the galaxies selected in the {\it F702W}-band image.
 We used the isophotal-aparture colors at the pixels selected in the
{\it F702W}-band image with a threshold of 26.7 mag
arcsec$^{-2}$. This was done
to suppress any contamination by very faint background objects
(especially blue ones) inside the apertures.  We also examined the
circular-aperture colors using a fixed aperture with a
$1.\hspace{-2pt}''5$ 
diameter, and obtained statistical results which are consistent
with those from the isophotal-aparture colors. 

 The half-light radius (equivalent radius) in the {\it F702W} band was measured by examining the photometric growth curve using the ellipsoidal apertures defined for each galaxy, the mean surface-brightness values quoted below are those within the ellipse of the half light.
 In order to quantify the morphology of the galaxies, we introduced asymmetry and concentration indices which were proposed in Abraham et al. (1994) and have been used in the subsequent studies on the morphology of faint galaxies (e.g., Abraham et al. 1996a,b). 
 The asymmetry index for a galaxy is defined as the flux ratio of the
residual counts in subtracting the 180$^{\circ}$-rotated image to the
total counts in the original image above the isophotal threshold. The
concentration index is the flux ratio between those within the inner
and the outer ellipses. The outer ellipse is defined so as to contain
all pixels above the surface-brightness threshold where the
ellipticity and position angle are determined by the intensity second
moment, which is calculated after re-centering with Gaussian
smoothing. The inner ellipse has the same ellipticity and position
angle, but the one third of the axis length.
 Note that the defined concentration index varies with the redshift,
even for galaxies with exactly the same light distribution. Therefore,
a direct comparison with the nearby morphological classification is difficult. These parameters should be used only for a relative classification among the detected objects.

\section{Properties of Galaxies in the Region of the Clusters}

 Figure 1 shows the differential number counts of galaxies detected on
the {\it F702W} image. We tentatively refer to the WFPC2 region within
40'' radius from the radio galaxy 3C 324 ($z = 1.21$) as the `region
of clusters' and the remaining region as the `field'. The excess of
the galaxy surface density is clearly seen at {\it F702W} $\sim$ 24--26 mag.
  For clusters at such a high redshift, it is difficult to decide the
membership of galaxies or to delineate the true cluster boundary. We
examined the significance of the galaxy surface density by changing
the radius of the region centered on 3C 324 and adopted a radius which
maximizes the significance of the surface-density excess. Dickinson
(1997a,b) shows that there are conspicuous peaks at $z = 1.15$ and $z
= 1.21$ and no other notable feature in the redshift distribution of the galaxies in the region within 30'' from 3C 324 (see figure 5 in Dickinson 1997b). There seems to be only a few cluster galaxies with measured redshifts in the region outside the 30'' radius. The area of our region of the clusters is 1.21 arcmin$^2$.

 In figure 2, we show the distribution of the colors of the galaxies
with {\it F702W}=24--26 mag. There is a strong excess of red galaxies
with ${\it F450W}-{\it F702W} >1$; there are almost no galaxies with
this color range in the adjacent field. The conspicuous feature in
this figure is, however, that the galaxies which contribute to the
surface-density excess in {\it F702W}-band image indeed have a wide
range of the colors. Especially, there is an strong peak of the
overdensity at the blue side, ${\it F450W}-{\it F702W}$ $\sim$ $-$1.0 -- $-$0.5.
 We plotted the expected ${\it F450W}-{\it F702W}$ color for galaxies
with various star-formation histories and ages observed at $z = 1.2$
in figure 3 using the evolutionary-synthesis code GISSEL96 (Bruzual,
Charlot 1993). Only those galaxies with on-going star-formation can
have such blue colors at $z = 1.2$. Thus, if these blue galaxies are
really members of the clusters at $z = 1.15$ or $z = 1.21$, a
significant fraction of the galaxies in the cluster region may have on-going star-formation activity, and their rest-frame UV emission is dominated by the star-formation component. 

 If the stellar population in a galaxy is dominated by old stars, only
a small amount of on-going star-formation can significantly affect its UV
emission. Since the ${\it F450W}-{\it F702W}$ color only samples the
rest-frame UV emission for galaxies at $z = 1.2$, the blue
${\it F450W}-{\it F702W}$ color cannot tell the scale of the
star-formation activity in a galaxy. Near-infrared data are essential
to evaluate such a scale. It is, however, important to note that a
fraction of the galaxies do not evolve {\it passively} in the clusters
at $z = 1.2$, but show evidence of on-going star-formation activity.

It is possible that the apparent excess of the blue objects is just
contamination by a foreground cluster or a group just superposed on
the line of sight. However, the blue ${\it F450W}-{\it F702W}$ color is not compatible with those of the passively-evolving early-type galaxies at intermediate redshifts, either. If we consider that most of the clusters observed at low and intermediate redshifts are dominated by old early-type galaxies, it is not likely that the blue galaxy excess is due to contamination by a foreground regular cluster.

 We also checked whether the excess of the blue galaxies is not just a
result of a statistical fluctuation, although the WFPC2 field may be
too small to estimate the true scatter of the field galaxy counts. We
compared the surface density of blue galaxies, with ${\it F450W}-{\it
F702W}$ $< -0.5$, of the randomly selected regions with 40'' radius
with that of the rest of the WFPC2 field, and found that most of the
regions with significant surface-density excess are localized near 3C
324. Figure~4 shows the distribution of objects with different colors
on the sky. The blue galaxies with ${\it F450W} - {\it F702W} < -0.5$
tend to locate near 3C 324, although the clustering is not as strong
as that of the red galaxies with ${\it F450W} - {\it F702W} > 1.0$.

 Next, we consider the size distribution of the galaxies in the region
of the clusters. Figure 5 shows the distribution of the apparent
half-light radius in the region of the clusters, which is very similar
to that in the adjacent field. It also shows that the galaxies which
contribute to the surface-density excess have various sizes. There is
no strong correlation between the galaxy size and the colors.

 Figure~6 shows the correlation between $R_{\rm hl}$ in physical scale
and $\langle$$\mu$$\rangle$, the mean surface brightness within $R_{\rm hl}$ for
the red and blue galaxies. We assumed that the galaxies are at
$z = 1.2$. The sequence of red galaxies has a slope which is
consistent with the Kormendy relation of the non-BCG elliptical
galaxies studied by Hoessel et al. (1987), while the blue galaxies
deviate from the relationship; these red galaxies must be early-type galaxies in the clusters. From the figure, we found that these
red galaxies may have $\sim 1$ mag positive evolution of luminosity at
3000 \AA\ in the rest frame or a factor of four negative evolution of
their half-light radius if we assume $q_0$ = 0.5. Because the latter possibility seems to be unreasonable, we prefer the former implication. The luminosity evolution as well as the observed color is consistent with those expected for passively evolving old galaxies. Similar analyses and results are presented in Dickinson (1997c).

 The quantitative morphology of the galaxies was also
investigated. The distributions of the asymmetry indices, log $A$, and
the concentration indices, log $C$, of the galaxies in the region of
the clusters and in the adjacent field are shown in  figure~7. The
distribution of these parameters of galaxies in the cluster field is
very wide, and there is no conspicuous difference between those in the
region of the clusters and in the field. Figure~8 shows the
surface-density of galaxies as a function of the morphological
parameter. We sorted the galaxies along an arbitrarily chosen line on
the log $C$ -- log $A$ diagram in order to put them in a sequence from
relatively asymmetrical and diffuse objects to relatively symmetrical
and concentrated objects. Again, an excess surface density can be seen for galaxies with a wide range of morphology.
 We then investigated the correlation between the morphology and the
color of the galaxies in the region of the clusters. Figure~9
demonstrate that the red galaxies  have a relatively symmetric and
large concentration. This is consistent with the idea that these red galaxies are indeed early-type galaxies which are progenitors of those in nearby rich clusters.  
 On the other hand, blue galaxies have a more asymmetric and diffuse
morphology than the red ones. The cluster members among these blue
galaxies are thus likely to have a different morphological type from the red quiescent (early-type) galaxies.
Dickinson (1997a) presented a montage of the galaxies at $z = 1.15$
and $z = 1.21$. Clearly, some of them have a diffuse or asymetric
morphology. We identified these galaxies on the {\it F702W}-band image
by comparing Figure 3 in Dickinson (1997a) with the data. They give
${\it F450W}-{\it F702W} \sim$ $-0.6$ -- 0.

 We also investigated the color distribution of the galaxies with
different morphology. As in figure 6, we sorted the galaxies along the
one-dimensional morphological sequence, and chose the `diffuse,
asymmetrical' galaxies (log $A-3\times$log $C$ $> 0.5$) and `concentrated,
symmetrical' galaxies(log $A-3\times$log $C$ $< 0.25$). We show the
color distribution of galaxies in these two samples in figures 10a and
10b. Clearly, some segregation in color distribution between the two samples exists.

\section{Conclusion and Discussion}

 We have shown that those galaxies which are responsible for the
surface-density excess on the {\it F702W}-band image have a wide range
of rest-UV color, size, and morphology. This implies that the UV
emission of galaxies in the clusters at $z \sim 1.2$ show different
characteristics from galaxy to galaxy. There is a notable
surface-density excess among the blue galaxies (${\it F450W}-{\it
F702W} <$ 1), which implies that there is a significant fraction of
galaxies which do not evolve passively and have some star-formation
activity at the observed epoch. These galaxies have a relatively diffuse or asymmetric morphology.
 These features are not likely to be due to contamination by
foreground objects, although there is a possibility that the wide
color distribution of galaxies may be due to a difference in the
properties between the cluster at 1.15 and that at 1.21. At such a
high redshift, the photometric properties of galaxies in clusters is
not expected to be as homogeneous as those seen in the near-by
universe. Thus, the galaxy properties discussed in this paper should
be considered as {\it average} ones of the two clusters at $z \sim 1.2$.

 The sample selected in {\it F702W}-band is rather biased to those
with UV-luminous objects, and it is not adequate to evaluate the blue
galaxy fraction to compare with those of clusters at an intermediate
redshift. In order to compare the integrated star-formation properties
of the clusters it is necessary to define a sample which is less
affected by star formation, thus, an NIR observation is essential. Instead, we here compare the properties of high- and intermediate-redshift galaxies with evidence of on-going star-formation.
 Dressler et al. (1999)  revealed the spectral properties of galaxies
in intermediate-redshift clusters. There are a class of ``e" or ``e(a),
e(b), and e(c)", where the galaxies have an [O\, {\footnotesize II}] emission line stronger
than EW([O {\footnotesize II}]) $>$ 5\AA . Galaxies with on-going star formation must be
classified to this category. These galaxies have $M_{V}$ $\sim$
$-22$ -- $-18.5$ ($H_0$ = 50 km s$^{-1}$ Mpc$^{-1}$, $q_0=0.5$) and
morphological type Sb -- Sd/Irr. There are also galaxies with
post-starburst components classified as``k+a" or ``a+k". These galaxies
also have $M_{V}$ = $-22$ -- $-19$, although their magnitude distribution
seems to be peaked at somewhat a brighter level than that of the
"e"-class galaxies. The morphological type of the ``k+a" or ``a+k" galaxies
is widely distributed from E to Sc. Assuming a typical color of
star-forming galaxy, $g-r \sim 0.5$, we evaluated that the absolute
magnitude at the rest-frame 3500 \AA\  $M_{\rm AB}$(3500) of these `active' galaxies is $-21$ -- $-17.5$.
 On the other hand, the blue galaxies in the clusters in the 3C 324
region with ${\it F450W}-{\it F702W} <$ 1 must be galaxies with
on-going star formation or those seen at $< 1$ Gyr after star
formation has stopped (figure 3). It is interesting that these blue
galaxies in 3C 324 clusters also have a relatively late morphological
type compared to the candidate of the progenitors of early-type
galaxies. Their absolute magnitude are $M_{\rm AB}$(3500) $\approx$ $-21$
-- $-19$ and in the range of active galaxies at an intermediate
redshift. A further systematic observation of galaxies in the clusters at high redshift will be necessary to reveal the true relationship among them.

%

\vspace{1pc}\par
 This work is based on observations with the NASA/ESA {\rm Hubble Space Telescope}, obtained at the Space Telescope Science Institute, operated by the Association of Universities for Research in Astronomy, Inc., from NASA contract NAS5-26555.


\section*{References}
\small

\re
Abraham R. G., Tanvir N. R., Santiago B. X., Ellis R. S., Glazebrook K., van den Bergh S. 1996a, MNRAS 279, L47 

\re
Abraham R. G., Valdes F. , Yee H. K. C., van den Bergh, S.  1994, ApJ 432, 75

\re
Abraham R. G., van den Bergh S., Glazebrook K., Ellis R. S., Santiago B. X., Surma P., Griffiths R. E. 1996b, ApJS 107, 1

\re
 Ben\i tez N, Broadhurst T., Rosati P., Courbin F., Squires G., Lidman C., Magain P. 1998, ApJ submitted (astro-ph/9812218)
 
\re
Bertin E., Arnouts S. 1996, A\&AS 117, 393 

\re
Bruzual A. G., Charlot S.  1993, ApJ 405, 538 

\re
Butcher H., Oemler A. Jr 1978, ApJ 219, 18 

\re
Butcher H., Oemler A. Jr 1984, ApJ 285, 426 

\re
Dickinson M. 1995, ASP Conf. Ser. 86, 283 

\re
Dickinson M. 1997a, in HST and the High Redshift Universe, ed
N.R. Tanvir, A. Arag\'on-Salamanca, J.V. Wall (World Scientific : Singapore) p207

\re
Dickinson M. 1997b, in The Early Universe with the VLT, ed J. Bergeron 
(Springer : Berlin) p274 

\re
Dickinson M. 1997c, in Galaxy Scaling Relations: Origins, Evolution and
Applications, ed L.N. da Costa, A. Renzini (Springer : Berlin) p215 

\re
Dressler A., Gunn J. E. 1982, ApJ 263, 533 

\re
Dressler A., Gunn J. E. 1983, ApJ 270, 7 

\re
Dressler A., Gunn J. E. 1992, ApJS 78, 1 

\re
Dressler A., Gunn J. E., Schneider D. P. 1985, ApJ 294, 70 

\re 
Dressler A., Oemler A. Jr, Butcher H. R., Gunn J. E. 1994, ApJ 
430, 107 

\re
Dressler A., Oemler A. Jr, Couch W. J., Smail I., Ellis R. S., Barger
A., Butcher H., Poggianti B. M. et al. 1997, ApJ 490, 577 

\re
Dressler A., Smail I., Poggianti B. M., Butcher H., Couch W. J., Ellis
R. S., Oemler A. Jr 1999, ApJS in press ({\it astro-ph}/9901263)

\re
{\rm HST Data Handbook, Ver3.1} 1998, (Space Telescope Science
Institute, Baltimore) http://www.stsci.edu/documents/data-handbook.html

\re
Hoessel J. G., Oegerle W. R., Schneider D. P. 1987, AJ 94, 1111 

\re
Kristian J., Sandage A., Katem, B. 1974, ApJ 191, 43 

\re
Hall P. B., Green R. F. 1998, ApJ 507, 558

\re
Poggianti B. M., Smail I., Dressler A., Couch W. J., Barger 
A. J., Butcher H., Ellis R. S., Oemler A. Jr 1999, ApJ in press ({\it astro-ph}/9901263)

\re
Rakos K. D., Maindl T. I., Schombert J. M. 1996, ApJ 466, 122 

\re
Rakos K. D., Schombert J. M. 1995, ApJ 439, 47 

\re
Smail I., Dickinson M.  1995, ApJ 455, L99 

\re
Spinrad H., Djorgovski S. 1984, ApJ 280, L9 

\re
Stanford S. A., Elston R., Eisenhardt P. R., Spinrad H., Stern D., Dey A. 1997, AJ 114, 2232

\re
Tanaka I., Yamada T., Arag\'on-Salamanca A., Kodama T., Miyaji 
T., Ohta K., Arimoto N. 1999, ApJ in press ({\it astro-ph}/9907437)

\re
Yamada T., Tanaka I., Arag\'on-Salamanca A., Kodama T., Ohta K., Arimoto N. 1997, ApJ 487, L125

\clearpage

\begin{figure*}
 \begin{center}
   \leavevmode  
   \epsfxsize=36pc
   \epsfbox{fig1.eps}
 \end{center}
\footnotesize Fig.\ 1.\ Magnitude distribution of the galaxies
detected in the {\it F702W}-band image taken with the WFPC2. The solid
line shows that of the region within 40'' radius from 3C 324 and the dotted line shows that of the outer region. The error-bar of the field count represents the square root of the number of galaxies detected in the region.
\end{figure*}

\begin{figure*}
 \begin{center}
   \leavevmode  
   \epsfxsize=36pc
   \epsfbox{fig2.eps}
 \end{center}
\footnotesize Fig.\ 2.\ Color distribution of the galaxies with 24 $<$ $R_{F702W}$ $<$
26. The meaning of the lines is similar to those in figure 1. Note
that the detection limit of the {\it F450W}-band image is about 28 mag.

\end{figure*}
\begin{figure*}
 \begin{center}
   \leavevmode  
   \epsfxsize=36pc
   \epsfbox{fig3.eps}
 \end{center}
\footnotesize Fig.\ 3.\ Expected color of the model galaxies with various types of
star-formation histories and age observed at $z = 1.2.$
\end{figure*}

\begin{figure*}
 \begin{center}
   \leavevmode  
   \epsfxsize=36pc
   \epsfbox{fig4.eps}
 \end{center}
\footnotesize Fig.\ 4.\ Distribution of galaxies with 24 $<$ ${\it F702W}$ $<$ 26 on the
WFPC2 image. The galaxies with different colors are shown by the different symbols.
\end{figure*}

\begin{figure*}
 \begin{center}
   \leavevmode  
   \epsfxsize=36pc
   \epsfbox{fig5.eps}
 \end{center}
\footnotesize Fig.\ 5.\ 
Distribution of the half-light radius of galaxies with 24 $<$
$R_{F702W}$ $<$ 26. The meaning of the lines is similar to those in figure 1.
\end{figure*}

\begin{figure*}
 \begin{center}
   \leavevmode  
   \epsfxsize=36pc
   \epsfbox{fig6.eps}
 \end{center}
\footnotesize Fig.\ 6.\ 
 Correlation between the half-light radius and the mean surface
brightness for red and blue galaxies with 24 $<$ ${\it F702W}$ $<$ 26. The solid line is a least-squares fit to the red-galaxies data and the dotted line shows the relation for non-BCG giant elliptical galaxies in clusters studied by Hoessel et al. (1987).
\end{figure*}

\begin{figure*}
 \begin{center}
   \leavevmode  
   \epsfxsize=36pc
   \epsfbox{fig7.eps}
 \end{center}
\footnotesize Fig.\ 7.\ 
{Distribution of the asymmetry and the concentration indices, log $A$
and log $C$ (see text for the definition), for ``cluster" and ``field"
galaxies with 24 $<$ ${\it F702W}$ $<$ 26.}
\end{figure*}

\begin{figure*}
 \begin{center}
   \leavevmode  
   \epsfxsize=36pc
   \epsfbox{fig8.eps}
 \end{center}
\footnotesize Fig.\ 8.\ 
Galaxy surface density plotted along an arbitrarily-defined galaxy
morphology sequence from ``asymmetric, diffuse" to ``symmetric,
concentrate". The galaxies with 24 $<$ ${\it F702W}$ $<$ 26 are treated and the meaning of the lines is similar to those in figure 1.
\end{figure*}

\begin{figure*}
 \begin{center}
   \leavevmode  
   \epsfxsize=36pc
   \epsfbox{fig9.eps}
 \end{center}
\footnotesize Fig.\ 9.\ 
{Log $A$ -- log $C$ plot for galaxies in the cluster region. Galaxies
with red color (${\it F450W}-{\it F702W} >$ 1) and blue one (${\it
F450W}-{\it F702W} <$ $-$0.5) are plotted with different symbols.}
\end{figure*}

\begin{figure*}
 \begin{center}
   \leavevmode  
   \epsfxsize=36pc
   \epsfbox{fig10.eps}
 \end{center}
\footnotesize Fig.\ 10.\ 
(a) Color distribution of the relatively ``asymmetric, diffuse" galaxies and (b) ``symmetric, concentrate" galaxies. The meaning of the lines is similar with those in figure 1.
\end{figure*}



\end{document}